\documentstyle[12pt]{article}
\newcommand{\ben}{\begin{equation}}{\rm }
\newcommand{\een}{\end{equation}}
\newcommand{\bea}{\begin{eqnarray}}{\bf }
\newcommand{\eea}{\end{eqnarray}}
\newcommand{\bear}{\begin{array}}
\newcommand{\enar}{\end{array}}
\newcommand{\bdm}{\begin{displaymath}}
\newcommand{\edm}{\end{displaymath}}
\newcommand{\no}{\nonumber}
 
\newcommand{\hf}{\frac{1}{2}}

\newcommand{\pa}{\partial}
\newcommand{\del}{\delta}
\newcommand{\eps}{\epsilon}
\newcommand{\al}{\alpha}

{\bf }

\newcommand{\g}{\gamma}

\newcommand{\ps}{\psi}
\newcommand{\psb}{\overline{\psi}}

\newcommand{\half}{\frac{1}{2}}
\newcommand{\p}{\phi}

\newcommand{\be}{\beta}

\newcommand{\tps}{\tilde{\psi}}

\newcommand{\sig}{\sigma}

\newcommand{\la}{\lambda}
\newcommand{\chb}{\bar{\chi}}
\newcommand{\Gh}{\hat{G}}
\newcommand{\Uh}{\hat{U}}
\newcommand{\Bh}{\hat{B}}
\newcommand{\Qh}{\hat{Q}}
\newcommand{\Th}{\hat{T}}
\newcommand{\Jh}{\hat{J}}

\begin{document}
\begin{flushright}
AS-ITP-2000-009\\
\vspace{1ex} 
September, 2000
\end{flushright}

\vspace{2ex}
\vspace{10ex}
\vspace{10ex}
\centerline{\Huge Superconformal Algebras in Light-cone Gauge}
\vspace{1ex}
\centerline{\Huge Quantization of String Theories on $AdS_3$ }
\vspace{2ex}
\vspace{1ex}
\vspace{1ex}
\vspace{1ex}
\vspace{2ex}
\vspace{2ex}
\vspace{3ex}
\centerline{\large {\sc Jian Jing}{\footnote{\tt e-mail: jingj@itp.ac.cn}} and  
{\sc Ming Yu }{\footnote{\tt e-mail: yum@itp.ac.cn}}}
\vspace{2ex}
\centerline{\it Institute of Theoretical Physics, Academia Sinica}
\vspace{1ex} 
\centerline{\it P.O.Box 2735, Beijing 100080, P.R.China}
\vspace{2ex}
\vspace{10ex}

\centerline{\large \it Abstract}
\vspace{1ex}
\begin{center}
\begin{minipage}{130mm}
{Motivated by superstring theories on $AdS_3$, we construct spacetime 
superconformal algebras (SCAs) living on the $AdS_3$ boundary in terms of the transversal physical degrees of freedom. 
The SCAs constructed are N=4 large, middle algebras, and $N=3$ algebra, corresponding to superstring theories on 
$AdS_3 \times S ^3 \times
S^3 \times S ^1$, $AdS_3 \times S ^3 \times T ^4 $ and $AdS_3 \times ( S^3\times S^3\times S^1)/Z_2$ backgrounds respectively.
}
\end{minipage}
\end{center}
\vfill
\newpage
 
 
\section{Introduction}
  String theory on $AdS$ space has been studied intensively, ~\cite{bofw, bars1, bars2,satoh} 
in particular, with the connection to  Maldacena's conjecture ~\cite {mal,andreev,boer,mal2,witten2,bars3,sugawara1,sugawara2, sugawara3,pe}. 
A precise statement on the $AdS/CFT$ correspondence has been given in ref. \cite{wit, pol}. 
In ref.\cite{gks,ks}, the spacetime
CFT living on the boundary of $AdS_3$ were studied and the $N=4$ superconformal
algebras on the spacetime boundary were constructed. Spacetime properties
of superstring theory on $AdS_3 \times S ^3 \times S^3 \times S ^1$ were studied in 
ref.\cite{efgt}, and the "large" $N=4$ algebras were constructed. Both of
those algebras were constructed in the covariant form. In ref.\cite{ymzb} the light-cone gauge quantization of string
theory on $AdS_3$ was given and shown to be equal to the covariant one, provided the center charge is 26 (Bosonic) or 15
(Fermionic). In the light-cone gauge, the spacetime superconformal algebra (SCA)  were 
constructed in terms
of the transversal physical degrees of freedom \cite {ymzb}. In this note we shall go along this 
line further. That is, we shall construct spacetime large, middle $N=4$ \cite{ali1,ali2}
and $N=3$ \cite{yis} SCAs in terms of the transversal physical degrees of freedom, corresponding to superstring theories on 
$AdS_3\times S^3 \times S^3 \times S^1$, $AdS_3\times S^3 \times T^4 $ and 
$AdS_3\times (S^3 \times S^3 \times S^1)/Z_2$ spacetime background respectively. \\
  This note is organized as follows. In section 2, we introduce our notations first, then we construct the large $N=4$
SCF algebra. In section 3, we get middle $N=4$ SCF algebra as well as $N=3$ SCF algebra from the large one by taking certain
limits. Some
remarks are given in the last section.\\

\section {Large $N=4$ spacetime SCFT}
  In this section we construct large $N=4$ SCF algebra, which corresponds to string propagating on $AdS_3 \times S ^3 \times S^3 
\times S ^1$ background. In Neveu-Schwarz-Ramond formalism, the worldsheet supersymmetry introduces ten worldsheet Majorana fermions, 
which however, transform as spacetime vectors. There are three worldsheet fermions
$\ps ^A$ , $"A"$ being the vector indices on the $SL(2)$ manifold, and similarly three fermions $\chi ^a$ from the first $SU(2)$ 
manifold, three fermions $\omega ^a$ from the second $SU(2)$ manifold, and a single fermion $\la$ from the $U(1)$ part. 
Our convention are almost the same as in ref. \cite {efgt} and \cite{ymzb}.
The OPEs among them are:
\bea
\ps ^A(z) \ps ^B(w) &=& \frac { \eta ^{AB}} {z-w} \no \\
\chi ^a(z) \chi ^b(w) &=& \omega ^a(z) \omega ^b(w) = \frac {\delta ^{ab}}{z-w} \\
\la(z)\la(w) &=& \frac {1} {z-w} \no
\eea
where $\eta ^{AB} = diag (+,+,-)$. \\	
In the $SL(2)\times SU(2)\times SU(2)\times U(1)$ super WZNW model, there are affine bosonic currents, $ j^A,
k ^i, m ^i$ with level $ k+2, k'-2, k''-2$, respectively. The OPEs among them are
\bea
j ^A(z) j ^B(w) &=& \frac { \frac {k+2}{2} \eta ^{AB}} {(z-w) ^2} + \frac {i\eta_{CD}
\eps^{ABC} j ^D} {z-w}  \no\\
k^i(z)k^j(w)&=& \frac { \frac {k'-2}{2} \delta ^{ij}}{(z-w) ^2}+ \frac {i\eps ^{ijk} k^k}{z-w} \no \\
m^i(z)m^j(w)&=& \frac { \frac {k''-2}{2} \delta ^{ij}}{(z-w) ^2}+ \frac {i\eps ^{ijk} m^k}{z-w} 
\no \\
\pa Y(z) \pa Y(w) &=& - \frac {1} {(z-w) ^2} 
\eea
The criticality of fermionic string, $c=15$, gives the following relation \cite{efgt}
\ben
\frac {1}{k} =\frac {1}{k'}+\frac {1}{k''}
\een
The $sl(2)$ current algebra can be constructed in terms of the three bosons as follows,
\bea
j^3&=&\be\gamma+\frac{\alpha_+}{2}\pa\p \no \\
j^+&=&\be{\gamma}^2+\alpha_+\g\pa\p+k\pa\g \label{rep}\\
j^-&=&\be \no
\eea 
where $\alpha_+ = \sqrt{2k}$. The OPEs of $\be, \gamma, \p$ are 
\bea
\p(z)\p(w)&=&-log(z-w)+\cdots\cdots\\
\be(z)\g(w)&=&\frac{1}{z-w}+\cdots\cdots \no
\eea
Here, $j^+=j^1+ij^2$, $j^-=j^1-ij^2$.\\
When the light-cone gauge \cite{ymzb} $\gamma = z^p$, and $\tilde\ps ^+ =0$ is imposed, only $\p$ and
$\tilde\ps^3$ (The OPE can be normalized as $\tilde\ps^3 (z) \tilde\ps^3(w)=\frac {1} {z-w}$) from 
$AdS_3$ part are dynamical ones. The bosonic part of the spacetime Virasoro generators are
\ben
L_n=\oint dz[-\half\pa\p\pa\p-(\frac{1}{\al_+}-\frac{\al_+}{2})
{\pa}^2\p+\frac {1}{k ^\prime} \sum k ^i k ^i + 
\frac {1}{k ^{\prime\prime}} \sum m^i m^i -\half \pa Y\pa Y-\frac{\Delta}
{z^2}]e^{nq}z^{np+1}
\een
Now we decompose the transversal physical degrees of freedom into the representations of the $su(2) \times 
su(2)$. The Liouville field $\phi$ and its super partner (${\tilde \ps}^3$) from the $AdS_3$ are in the 
$\bf{(0,0)}$. The $k^i$'s and their super partners ${\chi}^a$'s are in the $\bf{ (1,0)}$. Fields from 
$U(1)$ and its super partner are in the $\bf{(0,0)}$, and $m^i$'s and their super partners $\omega ^a$ in 
the $\bf{(0,1)}$. Since the spacetime supercurrents $G_{\al\be}$'s are spacetime spinors, and the 
worldsheet fermions are spacetime vectors, we first need to transform the later into spacetime spinors. 
This can be done by the usual bosonization procedure, as in ref. \cite {ymzb}\\
Define
\bea
\chi^3+\tps^3&=&e^{\p^1}\no\\
\chi^3-\tps^3&=&e^{-\p^1}\no\\
\chi^+&=&e^{\p^2}\no\\
\chi^-&=&e^{-\p^2}\\
\lambda^3+i\la^0&=&e^{\p^3}\no\\
\lambda^3-i\la^0&=&e^{-\p^3}\no\\
\lambda^1+i\la^2&=&e^{\p^4}\no\\
\lambda^1-i\la^2&=&e^{-\p^4}\no
\eea
Then we can construct the global $su(2)\times su(2)$ generators acting on these
fermions in terms  of the $\p^A$'s
\bea
j^+&=&\oint e^{\p^2}(e^{\p^1}+e^{-\p^1})\no\\
j^-&=&\oint (e^{\p^1}+e^{-\p^1})e^{-\p^2}\no\\
j^3&=&\oint \pa \p^2\\
k^+&=&\oint e^{\p^4}(e^{\p^3}+e^{-\p^3})\no\\
k^-&=&\oint (e^{\p^3}+e^{-\p^3})e^{-\p^4}\no\\
k^3&=&\oint \pa \p^4\no
\eea
Using the $\p^A$ fields and considering the representations of the $j^a$'s 
and $k^a$'s,
we can construct a set of worldsheet fermions which are also
spacetime spinors
\bea
\ps_1&=&e^{\hf(\p_1+\p_2+\p_3+\p_4)}\label{ps}\\
\psb_1&=&e^{\hf(-\p_1-\p_2-\p_3-\p_4)}\no\\
\ps_2&=&e^{\hf(-\p_1-\p_2+\p_3+\p_4)}\no\\
\psb_2&=&e^{\hf(\p_1+\p_2-\p_3-\p_4)}\no\\
\chi_1&=&e^{\hf(-\p_1+\p_2+\p_3-\p_4)}\no\\
\chb_1&=&e^{\hf(\p_1-\p_2-\p_3+\p_4)}\no\\
\chi_2&=&e^{\hf(\p_1-\p_2+\p_3-\p_4)}\no\\
\chb_2&=&e^{\hf(-\p_1+\p_2-\p_3+\p_4)}\no
\eea
where $\p_i$ satisfy $\p_i (z)\p_j (w) = \del_{ij} log (z-w)$, and the OPEs of ${\ps}_{\al},~\chi_{\al}$
are
\bea
{\ps}_{\al}(z){\psb}_{\be}(w)=\frac{{\del}_{\al \be}}{z-w}+ \cdots \cdots\\
{\chi}_{\al}(z){\chb}_{\be}(w)=\frac{{\del}_{\al \be}}{z-w}+ \cdots \cdots
\eea
and all others are regular. It is easy to see that both $\ps_i$'s and $\chi_i$'s are  
doublets of $j^i$
and $k^i$, $j^i$ acts from the left, and $k^i$  from the right \cite {ymzb}. The contributions to the spacetime energy-momentum 
tensor from fermions in terms 
of $\ps$'s, $\psb$'s and $\chi$'s, $\chb$'s can be expressed as 
\cite{ymzb}
\ben
T_F =\half(\pa{\ps}_1{\psb}_1-{\ps}_1\pa{\psb}_1+\pa{\ps}_2{\psb}_2-
{\ps}_2\pa{\psb}_2+\pa \chi_1 \chb_1 - \chi_1 \pa \chb_1 +\pa \chi_2 \chb_2 
- \chi_2 \pa \chb_2)
\een
where we have set $q=0,p=1$, so we can regard 
$$
T=-\half\pa\p\pa\p-(\frac{1}{\al_+}-\frac{\al_+}{2})
{\pa}^2\p+\frac {1}{k ^\prime} \sum k ^i k ^i + 
\frac {1}{k ^{\prime\prime}} \sum m^i m^i -\half \pa Y\pa Y + T _F
$$
as the energy-momentum tensor in spacetime.\\
Now, we have eight bosons and eight fermions. But this set of variables
are not convenient. For future convenience, we introduce two new variables to substitute for $\pa \phi$ and $\pa Y $\\
Define
\ben
J^0 = A \pa\p + B \pa Y, \qquad K^0 = C \pa \p + D \pa Y
\label{first}
\een
and
\ben
J^0(z)J^0(w) = K ^0(z) K ^0(w) = -\frac{1}{(z-w)^2}, \qquad J^0(z)K^0(w) = 0
\een
The coefficients $A,B,C,D$ satisfy
\ben
A^2+B^2=C^2+D^2=AD-BC=1, \qquad AC+BD=0
\een
we can choose the  solution
\ben
A=D=\sqrt\gamma, \qquad B=-C=-\sqrt{1-\gamma}
\een
where $\gamma =\frac{k''}{k'+k''}$. Of course both $J^0$ and $K^0$ belong to $\bf{ (0,0)}$.
So the energy-momentum tensor in spacetime can be rewritten as
\ben
T(z)=- \frac{1}{2} (J^0) ^2 - \frac{1}{2} (K^0) ^2 -(\frac{1}{\al_+} -
\frac{\al_+}{2}) {\sqrt \gamma} \partial J ^0 -(\frac{1}{\al_+} -\frac{\al_+}{2})
{\sqrt {1-\gamma}}\partial K^0 + \frac {1}{k ^\prime} \sum k ^i k ^i + 
\frac {1}{k ^{\prime\prime}} \sum m^i m^i + T _F
\label{second}
\een
Now we are ready to construct the large $N=4$ algebra, which consists of sixteen holomorphic currents 
\cite {ali1,efgt}. Apart from the energy-momentum tensor $T(z)$ and its four superpartners $G_{\al\be}$, there are seven  
currents $A^i, B^i,U$, which generate $su(2) \times su(2) \times u(1)$ subalgebra, (it arises from the $S ^3 \times
S^3 \times S^1$ background, and is the so-called "$R$ -symmetry" of the large $N=4$ algebra. The affine $su(2)\times 
su(2)\times u(1)$ algebra in the spacetime $N=4$ SCFT is lifted from the worldsheet.). Finally, there are four weight $\half$ 
fermionic generators $Q_{\al\be}$. They satisfy the following OPEs\\
\bea
T(z)T(w)& =& \frac {c/2} {(z-w) ^4} +\frac {2T(w)} {(z-w)^2} 
+ \frac {\pa T(w)}{z-w} \no \\
G_{\al\be}(z) G_{\al ' \be '}(w) &=& \eps_{\al\al'} \eps_{\be\be'} [\frac 
{2c/3} {(z-w)^3} + \frac {2T(w)} {z-w}] + {(\sig _i) _\al} ^\rho \eps_{\rho\al'}
\eps_{\be\be'}[\frac {4 \gamma A ^i}{(z-w) ^2} + \frac {2\gamma\pa A^i}{z-w}](w)
\no \\
&&+ {(\sig _i)_{\be'}} ^\rho \eps_{\rho\be} \eps_{\al\al'}[\frac {4(1-\gamma)B^i}
{(z-w)^2} + \frac {2(1-\gamma) \pa B^i} {z-w}] (w)\no \\
A^i (z)G_{\al\be}(w) &=& \half {(\sig _i) _\al} ^\rho [ \frac {G_{\rho\be}}
{z-w} -  \frac {2(1-\gamma) Q _{\rho\be}} { (z-w)^2}](w) \no \\
B^i (z)G_{\al\be}(w) &=& \half  {(\sig _i) _\be }^\rho[ \frac {G_{\al\rho}(w)}
{z-w} + \frac {2\gamma Q _{\al\rho}(w)} { (z-w)^2}]  \no \\
A^i (z) A^j (w)&=&\frac{k'/2 \delta^{ij}} {(z-w)^2} + \frac{i \eps^{ijk} A ^k(w)}
{z-w} \no \\
B^i (z) B^j (w)&=&\frac{k''/2 \delta^{ij}} {(z-w)^2} + \frac{i \eps^{ijk} B^k(w)}
{z-w} \no \\
Q_{\al\be}(z) G_{\al ' \be '}(w) &=&\frac {\eps_{\al\al'} \eps_{\be\be'}U(w)}{z-w}
+ \frac {1} {z-w}[ {(\sig _i) _\al} ^\rho \eps_{\rho\al'} \eps_{\be\be'} A^i
-{(\sig _i) _{\be'}} ^\rho  \eps_{\rho\be}\eps_{\al\al'}B^i](w) \no \\
A^i (z)Q_{\al\be}(w) &=& \half {(\sig _i) _\al} ^\rho  \frac {Q_{\rho\be}(w)}
{z-w} \no \\
B^i (z)Q_{\al\be}(w) &=& \half  \frac {{(\sig _i) _\be}^\rho Q_{\al\rho}(w) } {z-w}
 \no \\
U(z)G_{\al\be}(w)&=&\frac{Q_{\al\be}(w)}{(z-w)^2} \no \\
U(z)U(w) &=& -\frac {c}{12\gamma (1-\gamma)(z-w)^2} \no \\
Q_{\al\be}(z) Q_{\al ' \be '}(w) &=& -\frac{c}{12\gamma (1-\gamma)} 
\frac {\eps_{\al\al'} \eps_{\be\be'}}{z-w} \no \\
T(z) \Phi(w) &=& \frac { d _\Phi \Phi(w)} { (z-w)^2} + \frac {\pa \Phi(w)}{z-w} 
\label{last}
\eea
where $\sig _i =(\sig ^i)^*$, $\sig ^i$, $i=1,2,3$ is the  Pauli matrix,
\ben
\sig^1=\left(\begin{array}{cc}0&1\\1&0\end{array}\right),\ \
\sig^2=\left(\begin{array}{cc}0&-i\\i&0\end{array}\right),\ \
\sig^3=\left(\begin{array}{cc}1&0\\0&-1\end{array}\right)
\een 
$\Phi(z) = \{ G_{\al\be}, A^i, B^i, U, Q_{\al\be}\}$ and $d _\Phi = \{\frac{3}{2}, 1,1,1, \frac{1}{2}\}$ 
accordingly, $c=6k$.\\

We begin with the construction of  the $su(2) \times su(2) \times u(1)$ subalgebra. As in \cite{ymzb,yu}, the worldsheet current
algebra pertaining to the first $su(2)$ can be represented as
\ben
A^i=k^i+\half {\ps}_{\al} {\sig}^i_{\al \be} {\psb}_{\be}+\half {\chi}_{\al} {\sig}^i_{\al \be} {\chb}_{\be}\\
\een
the explicit expression is 
\bea
A ^+ &=& k ^+ + \ps_1 \psb_2 +\chi_1 \chb_2 \no \\
A ^3 &=& k ^3 + \half (\ps_1 \psb_1 -\ps_2 \psb_2) + \half (\chi_1 \chb_1 -
\chi_2 \chb_2) \no \\
A ^- &=& k ^- + \ps_2 \psb_1 + \chi_2 \chb _1 
\label{third}
\eea
which belong to $\bf(1,0)$. It is easy to check that 
\ben
A^i (z) A^j (w)=\frac{\delta^{ij}k'/2} {(z-w)^2} + \frac{i \eps^{ijk} A ^k(w)}{z-w} 
\label{a}
\een 
The second set of $su(2)$ currents can be constructed as follows
\bea
B ^+ &=& m ^+ +\chi_1 \chi_2 +\ps_1\ps_2 \no \\
B ^3 &=& m^3 + \half (\chi_1 \chb _1 + \chi_2 \chb_2 ) + \half (\ps_1\psb_1 +
\ps_2 \psb_2) \no \\
B ^- &=& m ^- +\chb_2 \chb_1 +\psb_2 \psb_1 
\label{fourth}
\eea
and they satisfy 
\ben
B^i (z) B^j (w)=\frac{\delta^{ij}k''/2} {(z-w)^2} + \frac{i \eps^{ijk} B^k(w)}{z-w} 
\label{b}
\een
which are in $\bf(0,1)$. The last $u(1)$ current can be constructed in a simple way
\ben
U(z)=-\sqrt {\frac {k'}{2}} J^0 +\sqrt {\frac {k''}{2} }K^0= \sqrt{\frac {k'+k''}{2}} \pa Y
\label{fifth}
\een
Here the coefficients are normalized to satisfy the large $N=4$ algebra. The spacetime affine $su(2) \times su(2) \times
u(1)$ subalgebra was given in ref. \cite {gks} 
\ben
\{A^a _n, B^a _n, U_n\} = \oint dz \{A^a(z), B^a(z), U(z)\} \gamma^n (z)
\een
In the light-cone gauge, they become simply
\ben
\{A^a _n,B^a _n, U_n \} = \oint dz \{A^a (z), B^a (z), U(z) \} e^{nq} z^{pn}
\een
Obviously, $A^a _n$'s and $B^a _n$'s form two affine $su(2)$ Lie algebras with level $k'(k'') _{st} = pk'(pk'')$. When 
$q=0, p=1$ is set, those expression will take rather simpler form, 
$$
\{ A^a _n, B^a _n, U_n \} = \oint dz \{A^a(z), B^a(z), U(z)\}z ^n
$$
So $A^a(z), B^a(z), U(z)$ can be regarded as affine Lie algebras on the spacetime's boundary.\\
Now only four fermionic generators $Q_{\al\be}$
and four supercurrents $G_{\al\be}$ are to be determined. They are both in $\bf (\half, \half)$, so we can write them as 
$2 \times 2$ matrices
\ben
Q=\left(\begin{array}{rr}Q_{11}&Q_{12}\\
Q_{21}&Q_{22}\end{array}\right), \ \
G=\left(\begin{array}{rr}G_{11}&G_{12}\\
G_{21}&G_{22}\end{array}\right), \ \
\een
Clearly, $A ^i$ belonging to $\bf (1,0)$ acts on the first index of $Q_{\al\be}(G_{\al\be})$ only, $B ^i$ in 
$\bf (0,1)$ acts on the second index. $Q_{11}$ is a combination of $\ps_1$ and $\chi_1$, 
$Q_{11}=A \ps_1 + B \chi_1$,
(we hope it will not lead to any confusions with (\ref {first})). $Q_{\al\be}$ can be determined by the $A^i, B^i$ action.
The coefficients $A$ and $B$ can be determined by using OPE, $ Q_{11}(z)
Q_{22}(w) = - {\frac {c} {12 \gamma (1-\gamma)}} {\frac {1} {z-w}}$. We can choose, 
$A=\sqrt {\frac {k'}{2}}$, $B=-\sqrt {\frac {k''}{2}}$ as the solution. So these four fermionic generators $Q_{\al\be}$ can 
be expressed explicitly 
\bea
Q_{11} = \sqrt {\frac {k'}{2}} \ps_1 - \sqrt {\frac {k''}{2} } \chi_1 ,\qquad
Q_{12} = \sqrt {\frac {k'}{2}} \psb_2 -\sqrt {\frac {k''}{2} } \chb_2 \no \\
Q_{21} = \sqrt {\frac {k'}{2}} \ps_2 - \sqrt {\frac {k''}{2} } \chi_2 ,\qquad
Q_{22} = -\sqrt {\frac {k'}{2}} \psb_1 +\sqrt {\frac {k''}{2} } \chb_1
\label{sixth}
\eea
Now, we construct the supercurrents. Consider all the  possibilities for constructing $G_{11}$ up to normalization
factors,
\ben
G_{11}=A[k^3\ps _1 + Bk ^+ \ps _2 +C\partial \ps _1+ D\ps _1 \psb_2 \ps _2 + EJ ^0 \ps _1]  
+ F[ m ^3 \chi _1 + Gm ^+ \chb _2 +H\partial \chi _1 + I\chi_1 \chb _2 \chi _2 + J K^0 \chi_1]
\een
where $A,B, \cdots , J$ (we also hope it will not lead to any ambiguities with (\ref{first})) are numeric coefficients to be 
determined. $A^+(z)G_{11}(w)$ regular requires $B=1$. $B^+(z)G_{11}(w)$ regular determines $G=1$. $G_{11}(z) G_{11}(w)$
regular determines $D=1, E=-\sqrt {\frac{k'}{2}}$ and $I=-1, J=-\sqrt {\frac{k''}{2}}$. The coefficients $A,F,C,H$  can be
determined by considering OPE between $G_{11}$ and $A^-$
as well as $G_{11}$ and  $B^-$. 
The other three supercurrets $G_{21},G_{12},G_{22}$ can be obtained by acting $A^-$ and $B^-$ on $G_{11}$..... In this
way, we get all supercurrents and we list them as follows
\bea
G_{11}&=&-\sqrt {\frac {2} {k ^\prime}} [k^3\ps _1 + k ^+ \ps _2 +(1- {\frac{c}{6}})
\partial \ps _1+ \ps _1 \psb_2 \ps _2 - \sqrt {\frac {k ^\prime}  {2}} J ^0 \ps _1] \no \\ 
&&-\sqrt {\frac {2} {k ^ {\prime \prime}}}[ m ^3 \chi _1 + m ^+ \chb _2 +
(1-\frac {c}{6}) \partial \chi _1 - \chi_1 \chb _2 \chi _2 - \sqrt 
{\frac {k ^{\prime\prime}}{2}} K^0 \chi_1] \no \\
G_{21}&=&-\sqrt {\frac {2} {k ^\prime}} [-k^3\ps _2 + k ^- \ps _1 +(1- {\frac{c}{6}})
\partial \ps _2+ \psb _1 \ps_1 \ps _2 - \sqrt {\frac {k ^\prime}{2}} J ^0 \ps _2] \no \\
 &&- \sqrt {\frac {2} {k ^ {\prime \prime}}}[ m ^3 \chi _2 - m ^+ \chb _1
 +(1-\frac {c}{6}) \partial \chi _2 - \chb_1 \chi _1 \chi _2 - \sqrt 
 {\frac {k ^{\prime\prime}}{2}} K^0 \chi_2] \no \\
G_{12}&=&\sqrt {\frac {2} {k ^\prime}} [-k^3\psb _2 + k ^+ \psb _1 -(1- {\frac{c}{6}})
\partial \psb _2+ \psb _1 \ps_1 \psb _2 + \sqrt {\frac {k ^\prime}  {2}} J ^0 \psb _2] \no \\ 
&&+ \sqrt {\frac {2} {k ^ {\prime \prime}}}[ m ^3 \chb _2 - m ^- \chi _1 -
(1-\frac {c}{6}) \partial \chb _2 - \chb_1 \chi _1 \chb _2 + \sqrt 
{\frac {k ^{\prime\prime}}{2}} K^0 \chb_2] \no \\
G_{22}&=&-\sqrt {\frac {2} {k ^\prime}} [k^3\psb _1 + k ^- \psb _2 -(1- {\frac{c}{6}})
\partial \psb _1+ \psb _2 \ps_2 \psb _1 + \sqrt {\frac {k ^\prime}  {2}} J ^0 \psb _1] \no \\
&&-\sqrt {\frac {2} {k ^ {\prime \prime}}}[ m ^3 \chb _1 + m ^- \chi _2 -
(1-\frac {c}{6}) \partial \chb _1 - \chb_2 \chi _2 \chb _1 + \sqrt 
{\frac {k ^{\prime\prime}}{2}} K^0 \chb_1]
\label{seventh}
\eea
where $c=6k=\frac {6k'k''}{k'+k''}$ \cite{gks,efgt}, as desired. Generators (\ref {second}), (\ref {third}), (\ref {fourth})
, (\ref {fifth}), (\ref {sixth}), (\ref{seventh}), form the complete basis of the large $N=4$ SCF algebra. It is a straightforward thing to check 
that they 
satisfy the OPEs (\ref {last})
 
\section{Middle $N=4$ and $N=3$ spacetime SCFT}
We shall construct the so-called middle $N=4$ and $N=3$ superconformal algebra on the spacetime boundary in this section. For the case
of middle $N=4$ superconformal algebra,
there are
sixteen holomorphic currents \cite {ali1,ali2, efgt} too, but the Kac-Moody subalgebra this time is $su(2) \times u(1) ^4$ 
(it arises from the
$S ^3 \times T ^4$ background ). \\
As pointed out in ref. \cite {efgt}, if either of $k'$ or $k''$, say, $k'' \to \infty$ (it amount to set $\gamma \to 1$), then in 
the OPE
$$ 
B^i(z)B^j(w) = \frac {k''/2 \delta^{ij}}{(z-w)^2} + \frac {i \eps^{ijk} B^k}{z-w}
$$
we can simply neglect the last term. It means that one of the $su(2)$ Kac-Moody algebra $B^i(z)$ is broken 
to $u(1)^3$ Kac-Moody $\Bh^i(z)$. From the spacetime point of view, it corresponds to the $AdS_3\times
S^3\times T^4$ background. The criticality requirement reduces to $k=k'$ and the center charge equals $6k$ now. We shall 
construct this middle $N=4$ algebra from the large one by means of the In$\ddot{\rm o}$n$\ddot{\rm o}$
-Winger contraction \cite {ali1}.\\
Make the following definitions
\bea
\Th(z)=\lim_{\gamma\to 1}T(z) ,\qquad \Uh^i(z)=\lim_{\gamma\to 1} \sqrt {1-\gamma}
B^i(z)\no\\
\Gh_{\al\be}(z)= \lim_{\gamma\to 1}G_{\al\be}(z), \qquad \Qh_{\al\be}(z) = \lim_{\gamma\to 1} \sqrt
{1-\gamma}Q_{\al\be}(z)\no\\
\Jh^i(z) = \lim_{\gamma\to 1}A^i(z),\qquad \Uh(z)= \lim_{\gamma\to 1} \sqrt {1-\gamma}
U(z)
\eea 
It seems that some of them are trivial and some of them have vanishing right hand sides. In fact neither of these 
observation are correct. The reason has been given in detail in ref.\cite {ali1}. First look at $\Uh^i(z) 
\Uh^j(w)$, we can easily get $\Uh^i(z) \Uh^j(w)=\frac {\del ^{ij}k'/2 }{(z-w)^2}=\frac {\del^{ij}c/12}{(z-w)^2}$, 
so we obtain $u(1)^3$ algebra 
$\Bh^i(z)$ which is broken from $B^i(z)$  as follows,
\ben
\Bh ^- = \Uh ^-, \qquad \Bh ^3= \Uh ^3, \qquad \Bh^+ = \Uh ^+
\een
and other generators (energy-momentum tensor $\Th$, supercurrent $\Gh_{ij}$ and four spin $\half$ fermions $\Qh_{ij}$) can be 
gotten in a similar way. We list these generators in the following,
\bea
\Th(z)&=&- \frac{1}{2} (J^0) ^2 - \frac{1}{2} (K^0) ^2 -(\frac{1}{\al_+} -
\frac{\al_+}{2}) \partial J ^0 + \frac {1}{k ^\prime} \sum k ^i k ^i + 
\frac {6}{c} \Uh ^i \Uh^i  + \Th _F \no \\
\Gh_{11}&=&-\sqrt {\frac {2} {k ^\prime}} [k^3\ps _1 + k ^+ \ps _2 +(1- {\frac{c}{6}})
\partial \ps _1+ \ps _1 \psb_2 \ps _2 - \sqrt {\frac {k ^\prime}  {2}} J ^0 \ps _1] \no \\ 
&&+(K ^0 \chi_1 - \sqrt {\frac {12}{c}} \Uh ^3 \chi_1 - \sqrt {\frac {12}{c}} \Uh^+
\chb_2) \no \\
\Gh_{21}&=&-\sqrt {\frac {2} {k ^\prime}} [-k^3\ps _2 + k ^- \ps _1 +
(1- {\frac{c}{6}}) \partial \ps _2+ \psb _1 \ps_1 \ps _2 - \sqrt 
{\frac {k ^\prime}{2}} J ^0 \ps _2] \no \\
 &&+ (K ^0 \chi_2 - \sqrt {\frac {12}{c}} \Uh ^3 \chi_2 + \sqrt {\frac {12}{c}} \Uh^+
\chb_1) \no \\
\Gh_{12}&=&\sqrt {\frac {2} {k ^\prime}} [-k^3\psb _2 + k ^+ \psb _1 -(1- {\frac{c}{6}})
\partial \psb _2+ \psb _1 \ps_1 \psb _2 + \sqrt {\frac {k ^\prime}  {2}} J ^0 \psb _2] \no \\
&&+ (K ^0 \chb_2 + \sqrt {\frac {12}{c}} \Uh ^3 \chb_2 - \sqrt {\frac {12}{c}} \Uh^-
\chi_1) \no \\
\Gh_{22}&=&-\sqrt {\frac {2} {k ^\prime}} [k^3\psb _1 + k ^- \psb _2 -(1- {\frac{c}{6}})
\partial \psb _1+ \psb _2 \ps_2 \psb _1 + \sqrt {\frac {k ^\prime}  {2}} J ^0 \psb _1] \no \\
&&-(K ^0 \chb_1 + \sqrt {\frac {12}{c}} \Uh ^3 \chb_1 + \sqrt {\frac {12}{c}} \Uh^-
\chi_2) \no \\
\Uh(z)&=&\sqrt {\frac {c}{12}} K ^0 \no \\
\Bh ^- &=& \Uh ^-, \qquad \Bh ^3= \Uh ^3, \qquad \Bh^+ = \Uh ^+ \no \\
\Qh_{11} &=& -\sqrt {\frac {c}{12}} \chi_1, \qquad \Qh_{12} = -\sqrt {\frac {c}{12}} \chb_2 \no \\
\Qh_{21} &=& -\sqrt {\frac {c}{12}} \chi_2, \qquad \Qh_{22} = \sqrt {\frac {c}{12}} \chb_1 \no \\
\Jh^+ &=& k ^+ + \ps_1 \psb_2 +\chi_1 \chb_2 \no \\
\Jh^3 &=& k ^3 + \half (\ps_1 \psb_1 -\ps_2 \psb_2) + \half (\chi_1 \chb_1 -
\chi_2 \chb_2) \no \\
\Jh^- &=& k ^- + \ps_2 \psb_1 + \chi_2 \chb _1 
\eea

where  
$$
\Th _F = T_F=\half(\pa{\ps}_1{\psb}_1-{\ps}_1\pa{\psb}_1+\pa{\ps}_2{\psb}_2-
{\ps}_2\pa{\psb}_2+\pa \chi_1 \chb_1 - \chi_1 \pa \chb_1 +\pa \chi_2 \chb_2 
- \chi_2 \pa \chb_2)
$$
are unchanged.

They satisfy the following middle $N=4$ superconformal algebra
\bea
\Th(z)\Th(w)& =& \frac {c/2} {(z-w) ^4} +\frac {2\Th(w)} {(z-w)^2} 
+ \frac {\pa \Th(w)}{z-w} \no \\
\Gh_{\al\be}(z) \Gh_{\al ' \be '}(w) &=& \eps_{\al\al'} \eps_{\be\be'} [\frac 
{2c/3} {(z-w)^3} + \frac {2\Th(w)} {z-w}] + {(\sig _i) _\al} ^\rho \eps_{\rho\al'}
\eps_{\be\be'}[\frac {4 \Jh ^i}{(z-w) ^2} + \frac {2\pa \Jh^i}{z-w}](w) \no\\
\Jh^i (z)\Gh_{\al\be}(w) &=& \half {(\sig _i) _\al} ^\rho \frac {\Gh_{\rho\be}(w)}
{z-w} \no \\
\Bh^i (z)G_{\al\be}(w) &=& \frac {{(\sig _i) _\be}^\rho \Qh_{\al\rho}(w) }
{(z-w)^2} \no \\
\Uh(z) \Gh _{\al\be}&=& \frac {\Qh_{\al\be}(w)}{(z-w)^2} \no \\
\Jh^i (z) \Jh^j (w)&=&\frac{k'/2 \delta^{ij}} {(z-w)^2} + \frac{i \eps^{ijk} \Jh ^k(w)}
{z-w} \no \\
\Bh^i (z) \Bh^j (w)&=&\frac{c \delta^{ij}} {12(z-w)^2}  \no \\
\Qh_{\al\be}(z) \Gh_{\al ' \be '}(w) &=&\frac {\eps_{\al\al'} \eps_{\be\be'}\Uh(w)}{z-w}
- \frac {{(\sig _i) _{\be'}}^\rho \eps_{\rho\be} \eps_{\al\al'}\Bh^i(w) } {z-w}\no \\
\Qh_{\al\be}(z)\Qh_{\al'\be'}(w)&=&-\frac{c/12 \eps_{\al\al'} \eps_{\be\be'}}
{z-w} \no \\
\Jh^i (z)\Qh_{\al\be}(w) &=& \half {(\bar \sig _i) _\al} ^\rho  \frac {\Qh_{\rho\be}(w)}
{z-w} \no \\
\Uh(z)\Uh(w) &=&-\frac{c/12}{(z-w)^2} \no \\
\Uh ^i(z) \Uh(w) ^j &=& \frac {c/12\delta^{ij}}{(z-w)^2} \no \\
\Th(z) \hat{\Phi}(w) &=& \frac { d _{\hat\Phi} {\hat\Phi}(w)} { (z-w)^2} + 
\frac {\pa {\hat\Phi}(w)}{z-w} 
\eea
where  $\hat \Phi = \{ \Gh _{\al\be}, \Bh ^i, \Uh, \Jh ^i, \Qh _{\al\be} \}$, accordingly, $d _{\hat\Phi}$ equal to $ \{ \frac {3}{2}, 1,1,1,
\frac {1}{2} \}$, and $c=6k$, as mentioned above.\\

Recently, string theory on the $AdS_3\times( S^3\times S^3\times S^1)/Z_2$ was 
investigated in ref.\cite{yis}, and the spacetime $N=3$ superconformal theories
was constructed. We shall construct the same algebra in terms of the transversal
physical degrees of freedom.\\
We set $k'=k''$ in our large $N=4$ algebra. Once this setting is performed, there is an automorphism in our 
large $N=4$
algebra, namely, a $Z_2$ action\cite {yis}, 
$$(A^i(z), B^i(z), Y(z)) \to (B^i(z), A^i(z), -Y(z))$$
We consider the eigenstates of the $Z_2$ action. Those with eigenvalue $+1$ are 
invariant under $Z_2$ and will form a subalgebra of the large $N=4$ algebra. It can be checked that the diagonal part of the $su(2) 
\times su(2)$ algebra is invariant under $Z_2$ and the isometry $su(2) \times su(2)$ of the large $N=4$ is reduced to the diagonal
$su(2)$.\\
The diagonal $su(2)$ is invariant under $Z_2$, so we can construct our affine
Lie algebra uniquely by summing those two $su(2)$s
$$ J^i(z) = A^i(z)+B^i(z)$$
Supercharges $G_{ij}$ belong to the $\bf{ (\hf,\hf)}$ representation of $su(2)
\times su(2)$, and they are decomposed in to $\bf 1 \oplus \bf 0$ under the diagonal $su(2)$.
Now we shall explain that only the triplet is $Z_2$ invariant.\\
Under the $Z_2$ action, two $su(2)$s are interchanged and at the same time $Y$
is reflected. In our free field realization, 
this correspond to $A^i$ and $B^i$ interchanged, and $U$ reflected simultaneously according to (\ref{fifth}). Supercurrents and four fermions transform as 

\ben
Q=\left(\begin{array}{rr}Q_{11}&Q_{12}\\
Q_{21}&Q_{22}\end{array}\right) \to 
-\left(\begin{array}{rr}Q_{11}&Q_{21}\\
Q_{12}&Q_{22}\end{array}\right),
G=\left(\begin{array}{rr}G_{11}&G_{12}\\
G_{21}&G_{22}\end{array}\right) \to 
\left(\begin{array}{rr}G_{11}&G_{21}\\
G_{12}&G_{22}\end{array}\right)
\een
It is not very difficult to check that after $Z_2$ action, those set of generators still 
satisfy OPEs (\ref{last}), except $A^i, B^i$ interchanged, and the index of supercurrents also exchanged. So only triplet is invariant under
the $Z_2$ action. Closure of this algebra need a spin $\half$ fermionic current $\Psi$. We list our $N=3$ algebra 
generators in the following
\bea
T(z)&=&- \frac{1}{2} (J^0) ^2 - \frac{1}{2} (K^0) ^2 -\frac{\sqrt {2}}{2}
(\frac{1}{\al_+} -
\frac{\al_+}{2})  \partial J ^0 -\frac{\sqrt{2}}{2} (\frac{1}{\al_+} -
\frac{\al_+}{2})
\partial K^0 \no \\
&&+ \frac {1}{k ^\prime} \sum k ^i k ^i + 
\frac {1}{k ^{\prime\prime}} \sum m^i m^i + T _F \no \\
T_F &=&\half(\pa{\ps}_1{\psb}_1-{\ps}_1\pa{\psb}_1+\pa{\ps}_2{\psb}_2-
{\ps}_2\pa{\psb}_2+\pa \chi_1 \chb_1 - \chi_1 \pa \chb_1 +\pa \chi_2 \chb_2 
- \chi_2 \pa \chb_2) \no \\
J^+ &=& A^+ + B^+, \qquad J^3 = A^3 + B^3, \qquad J^- = A^- + B^- \no \\
G^+ &=& i\sqrt{2} G _{11},\qquad G^3 =-i \frac {\sqrt{2}}{2} (G_{12}+G_{21}),
\qquad G^- = -i\sqrt{2} G _{22} \no \\
\Psi&=&-i \frac {\sqrt{2}}{2}(Q_{21}-Q_{12})
\eea
where  the coefficients are normalized to satisfy the following OPEs
\bea
T(z)T(w)& =& \frac {c/2} {(z-w) ^4} +\frac {2T(w)} {(z-w)^2} 
+ \frac {\pa T(w)}{z-w} \no \\
J^i (z) J^j (w)&=&\frac{k' \delta^{ij}} {(z-w)^2} + \frac{i \eps^{ijk} J^ k(w)}
{z-w} \no \\
J^i (z) G^j (w)&=&\frac{i \eps^{ijk} G^ k(w)}{z-w} +\frac{\delta^{ij}\Psi(w)}
 {(z-w)^2} \no\\
G^i(z) G ^j (w)&=&\frac{2c/3 \delta^{ij}}{(z-w)^3} + \frac {2i \eps^{ijk}J ^k(w)}
{(z-w)^2 } +\frac {i \eps^{ijk} \pa J^k} {z-w} + \frac {2 \delta ^{ij} T(w) }{z-w}\no \\
\Psi(z) \Psi(w)&=&\frac{k'}{z-w} \no \\
T(z) \hat\Phi(w) &=& \frac { d _{\hat\Phi} {\hat\Phi(w)}} { (z-w)^2} + \frac {\pa \hat\Phi(w)}{z-w}  
\eea
where $\hat\Phi= \{ J^i, G ^i, \Psi$ \}, $d _{\hat\Phi}= \{1, \frac {3}{2}, \frac{1}{2} \}$, and the center charge now is $ c=6k=3k'=3k''$ 
(remember $k=\frac {k'} {2} =\frac {k''} {2}$ in this case).

\section {Conclusions and Remarks}
To have a better understanding of the $AdS/CFT$ correspondence, it is essential to construct the spacetime boundary CFT explicitly. In this
note we have realized the spacetime SCF algebras explicitly in terms of transversal degrees of freedom, corresponding to strings propagating on 
$AdS_3 \times S ^3 \times
S^3 \times S ^1$, $AdS_3 \times S ^3 \times T ^4 $ and $AdS_3 \times ( S^3\times S^3\times S^1)/Z_2$ respectively. This is a modest step towards to full 
construction of the spacetime CFT living on the $AdS_3$ boundary. \\
As in the case of the large $N=4$ SCF algebra \cite {efgt}, we may consider the spacetime SCFT as some twisted version of $p$ copies of nonlinear
sigma model, each with center charge $6k$. It remains to check the correspondence between the string theory in the bulk and the CFT on the boundary.\\
A manifest similarity between the two theory is that the DDF state \cite {gsw,sugawara4}on the string theory side resembles very much to the conformal algebra
on the boundary CFT side, except that the energy momentum  tensor on the each side differ by a improved term. Does that means some duality between
the two CFT? It deserves further clarification.

\section*{Acknowledgment}
One of the authors (J. Jing) would like to appreciate Dr. Bihn Zhou  and Dr. Shi-hua Zhao for their constructive discussions and 
enthusiastic helps.


\end{document}